\documentclass[twocolumn, showpacs,superscriptaddress,nofootinbib,floatfix, amsfonts, aps]{revtex4}
\usepackage{amsmath}
\usepackage{bm}
\usepackage{graphicx}
\usepackage{mathrsfs}
\usepackage{footmisc}
\usepackage{multirow, ulem}
\usepackage{graphicx}
\usepackage{booktabs, ctable}

\begin{document}

\title{Bottomonium Continuous 
Production from Unequilibrium Bottom Quarks in Ultrarelativistic Heavy Ion Collisions}
\author{Baoyi Chen}
\affiliation{Department of Physics, Tianjin University, Tianjin 300350, China }
\author{Jiaxing Zhao}
\affiliation{Department of Physics and Collaborative Innovation Center of Quantum Matter, 
Tsinghua University, Beijing 100084, China}

\date{\today}

\begin{abstract}
We employ the Langevin equation and Wigner function to describe the bottom equark dynamical evolutions 
and their formation into a bound state in the expanding Quark Gluon Plasma (QGP). 
The additional suppressions from parton inelastic scatterings 
are supplemented in the regenerated bottomonium. Hot medium 
modifications on $\Upsilon(1S)$ properties are studied consistently by taking the 
bottomonium potential to be the color-screened potential from Lattice results, 
which affects both $\Upsilon(1S)$ 
regeneration and dissociation rates. Finally, we calculated the 
$\Upsilon(1S)$ nuclear modification factor $R_{AA}^{\rm rege}$ from bottom quark combination 
with different diffusion coefficients in Langevin equation, 
representing different thermalization of bottom quarks. In the central Pb-Pb collisions (b=0) 
at $\sqrt{s_{NN}}=5.02$ TeV,  
we find a non-negligible $\Upsilon(1S)$ regeneration, and it is small 
in the minimum bias centrality. The connections between 
bottomonium regeneration and 
bottom quark energy loss in the heavy ion collisions 
are also discussed. 

\end{abstract}
\maketitle


\section{Introduction}

Since charmonium was proposed as a probe for the existence of the deconfined matter 
called ``Quark-Gluon Plasma'' (QGP)~\cite{Matsui:1986dk}, its production mechanisms 
has been widely studied based on coalescence 
model~\cite{Thews:2000rj,Greco:2003vf,Andronic:2003zv} 
and transport models~\cite{Grandchamp:2002wp,Yan:2006ve,Chen:2013wmr} 
in nucleus-nucleus collisions. The nuclear modification factor $R_{AA}$ 
is a measurement of cold and hot medium suppressions on quarkonium yields. 
The cold nuclear matter effects include the nuclear absorption~\cite{Vogt:1999cu}, 
Cronin and shadowing effects~\cite{Gavin:1988tw,Eskola:1998df,Vogt:2010aa}. The first 
one is negligible at LHC colliding energies due to strong Lorentz dilation, 
where ``spectator'' nucleons already move out of the colliding region before the formation of 
a quarkonium eigenstate. Cronin effect will shift the momentum 
distribution of primordially produced hidden- and 
open-charm(or bottom) states~\cite{Zhu:2004nw,Chen:2015ona}. 
This can be included by a proper modification of their transverse momentum 
distributions in pp collisions~\cite{Yu:2017pot,Chen:2015iga}. 
Shadowing effect is weak at RHIC colliding energies, but important at 
the LHC colliding energies. All the cold nuclear matter effects can be included in the heavy flavor 
initial distributions prior to the hot medium effects. In nucleus-nucleus collisions at 
$\sqrt{s_{NN}}=2.76$ TeV and 5.02 TeV, 
regeneration from charm and anti-charm quarks is widely believed to dominate the 
prompt charmonium yields~\cite{Zhao:2011cv,Liu:2009wza}. This is supported by 
the enhancement of $J/\psi$ $R_{AA}$ and the suppression of $J/\psi$ mean 
transverse momentum square 
$\langle p_T^2\rangle_{J/\psi}$ observed at 2.76 TeV and 5.02 TeV: 
the regenerated $J/\psi$s from 
thermalized charm quarks carry small momentum compared with the primordially produced ones, 
this will pull down the $\langle p_T^2\rangle_{J/\psi}$ of final prompt $J/\psi$ in 
nucleus-nucleus collisions~\cite{Zhou:2014kka}.

However for bottomonium, the situation seems not so clear. 
Transport model calculations suggest a non-negligible bottomonium regeneration in 
$\sqrt{s_{NN}}=5.02$ TeV Pb-Pb collisions~\cite{Zhao:2012gc}. 
Also, experimental data hinted a stronger bottomonium 
regeneration at 5.02 TeV compared with 2.76 TeV, but within its large uncertainty 
which prevents solid conclusions~\cite{Abelev:2014nua,Fronze:2016gsr}. 
Considering that heavy quark mass is very large, it takes some time to reach kinetic 
thermalization~\cite{vanHees:2004gq,vanHees:2007me} in the fast cooling 
QGP with an initial temperature $\sim 500$ MeV in AA collisions. 
Non-thermalization of bottom quark momentum distribution will suppress 
the combination probability of $b$ and $\bar b$ quarks in QGP. However, the ratio of 
hidden- to open-bottom states is at the order of $0.1\%$ which is smaller than the charm flavor. 
This may make the yield from ($b+\bar b\rightarrow \Upsilon(1S) +g$, 
$b+\bar b +\zeta \rightarrow \Upsilon(1S)+\zeta$ 
with $\zeta= g,q,\bar q$) 
not negligible compared with 
primordially produced $\Upsilon(1S)$. 
It is very interesting to develop a realistic model to consider the dynamical 
evolutions of bottom quarks, bottomonium regeneration process and the following hot medium 
suppression after they are regenerated. We employ the ``Langevin equation + Wigner function 
+ (gluon, quasi-free) dissociations'' for bottom quark and bottomonium 
evolutions in heavy ion collisions. Furthermore, we consider the hot medium modifications on bottomonium 
properties at finite temperature by taking the color-screened 
heavy quark potential (extracted from Lattice 
free energy $F(r,T)$~\cite{Digal:2005ht}). With color-screened heavy quark potential in time independent 
Schr\"odinger equation, we obtain the mean radius and binding energy of $\Upsilon(1S)$ at different 
temperatures. Each of them will be used in Wigner function (regeneration rate) and (gluon, quasi-free) 
dissociation rates. 

Our paper is organized as follows. In Section II, we introduce the Langevin equation and Wigner 
function for heavy quark dynamical evolutions and their combination. 
The hot medium 
modifications on $\Upsilon(1S)$ properties (such as the mean radius and binding energy) are also studied 
based on potential model. In Section III, we introduce the hydrodynamic equations for QGP expansion 
in nucleus-nucleus collisions. The relevant inputs of heavy quarks 
are presented in Section IV. In Section V, we give the 
$\Upsilon(1S)$ nuclear modification factor $R_{AA}^{\rm rege}$ 
from the combination of $b$ and $\bar b$ quarks in QGP. 
Different coupling strength between heavy quarks and QGP (controlling the heavy quark 
thermalization) are studied in the $\Upsilon(1S)$ regeneration. We summarize the work in Section VI.

\section{dynamical evolutions of heavy quarks}
\subsection{Langevin Equation for Heavy Quark Diffusion}
The heavy quark diffusion in Quark Gluon Plasma can be treated as a Brownian motion, 
which is widely studied with Langevin equation~\cite{He:2011qa,Cao:2013ita}. 
During the evolution of heavy quarks, they can combine into a 
quarkonium~\cite{Thews:2000rj,Greco:2003vf,Andronic:2007bi,Du:2015wha}. 
The probability of the combination of heavy quark $Q$ and $\bar Q$ depends on their 
distribuitons in phase space and also 
the properties of the produced quarkonium at 
finite temperature~\cite{Morita:2009qk,Chen:2012gg}. Instead of dealing 
with heavy quark distributions, we 
employ the Langevin equation to simulate $Q$ and $\bar Q$ evolutions in the hot medium. 
Combination process ($Q+\bar Q\rightarrow (Q\bar Q)_{\rm bound} +g$, 
$Q+\bar Q+\zeta \rightarrow (Q\bar Q)_{\rm bound} +\zeta$ 
with $\zeta=g,q,\bar q$) can be included through the Wigner function 
$W_{Q\bar Q\rightarrow \Upsilon}$~\cite{Greco:2003vf}. The Langevin equation is written as  
\begin{align}
{d\vec p\over dt}=-\vec \eta_D(p)\vec p+\vec \xi 
\label{fun-LG}
\end{align}
where $\vec \eta_D(p)$ and $\vec \xi$ are the drag force and the noise of the hot medium on 
heavy quarks. $\vec \xi$ satisfies the correlation relation
\begin{align}
\langle \xi^{i}(t)\xi^{j}(t^\prime)\rangle =\kappa \delta ^{ij}\delta(t-t^\prime)
\end{align}
$\kappa$ is the diffusion coefficient of heavy quarks in momentum space, which is connected with 
spatial diffusion coefficient $D$ by $\kappa=2T^2/D$. 
The drag force in Langevin equation 
can then be determined by the 
fluctuation-dissipation relation~\cite{Cao:2012jt}
\begin{align}
\label{eq-drag}
\eta_D(p)={\kappa\over 2TE}
\end{align}
$T$ is the temperature of the bulk medium and $E=\sqrt{m_Q^2+|\vec p|^2}$ 
is the heavy quark energy. 

In order to numerically solve the Langevin equation for heavy quark diffusions in QGP, 
it is discretized as below 
\begin{align}
\label{eq-numLan}
&\vec p(t+\Delta t) =\vec p(t)-\eta_D(p)\vec p \Delta t+\vec \xi\Delta t \\
\label{eq-LanR}
&\vec X_Q(t+\Delta t) = \vec X_Q(t) +{\vec p\over E}\Delta t \\
&\langle \xi^{i}(t)\xi^j(t-n\Delta t)\rangle ={\kappa \over \Delta t}\delta ^{ij}\delta^{0n}  
\label{fun-LD}
\end{align}
As long as the time step $\Delta t$ for numerical evolutions is small enough, one can assume 
free motions for heavy quarks with a constant velocity 
$\vec v_Q=\vec p/E$ during this time step, and update the 
heavy quark momentum by Eq.(\ref{eq-numLan}) at the end of each $\Delta t$ 
due to hot medium effects. The medium-induced 
radiative energy loss~\cite{Wang:1991xy,Baier:1996kr} 
and parton elastic collisions~\cite{Qin:2007rn} of heavy quarks can be included in the 
terms of the drag force 
$\vec \eta_D(p)$ and the noise $\vec \xi$. 
$\xi^i(t)_{i=1,2,3}$ in Eq.(\ref{fun-LD}) is sampled randomly based on a Gaussian function with the 
width $\sqrt{\kappa/\Delta t}$. 

The initial transverse momentum distribution as an input of Eq.(\ref{eq-numLan}) is obtained from 
PYTHIA simulations. As the initial energy density of QGP changes with coordinates, the heavy quark 
initial distribution in QGP affects their evolutions, which 
will finally affect the heavy quark thermalization degree and quarkonioum regeneration. 
Heavy quark pairs are produced from parton hard scatterings, their density (within rapidity region 
$\Delta y$) is proportional to the number of binary collisions,  
\begin{align}
\label{eq-den}
{dN_{\rm PbPb}^{Q\bar Q}\over d\vec x_T}=\sigma_{pp}^{Q\bar Q}(\Delta y)
\times T_{\rm Pb}(\vec x_T-{\vec b\over 2})T_{\rm Pb}(\vec x_T +{\vec b\over 2})
\end{align}
where $\sigma_{pp}^{Q\bar Q}(\Delta y)$ is the heavy quark production 
cross section in proton-proton collisions 
within rapidity $\Delta y$. 
$T_{\rm Pb}(\vec x_T)=\int dz \rho_{\rm Pb}(\vec x_T, z)$ is the 
thickness function of lead. Nucleus density $\rho_{\rm Pb}(\vec x_T, z)$ is taken to be 
the Woods-Saxon distribution. 
When colliding energy is at the order of $\sim$TeV, theoretical and experimental 
studies indicate a strong nucleus (anti-)shadowing effect 
on heavy quark (quarkonium) production~\cite{Vogt:2010aa}. 
Furthermore, this effect depends on the nucleon density, 
which gives different modification on 
heavy quark production at different positions of the nucleus. We employ a theoretical model (EPS09 NLO) to 
obtain a shadowing factor $r_S({\vec x}_T, p_T,y)$. The initial distribution of heavy quarks (quarkonium) 
in Pb-Pb collisions is then taken as ${dN_{\rm PbPb}^{Q\bar Q}\over d\vec x_T}\times r_S({\vec x}_T,p_T, y)$. 

\subsection{Heavy Quark Recombination Process with Wigner Function}
In the nucleus collisions, Heavy quark pairs are produced and evolve inside the QGP as a Brownian 
motion. During QGP evolutions, $Q$ and $\bar Q$ can meet each other and 
combine into a bound state, which 
may survive from the hot medium due to its large binding energy. If $Q$ and $\bar Q$ can reach 
kinetic equilibrium, their relative momentum ($\vec p_Q-\vec p_{\bar Q}$) and relative 
distance ($\vec X_Q-\vec X_{\bar Q}$) will be small, which enhances the combination probability of $Q$ and 
$\bar Q$~\cite{He:2014tga}. 
The discussion about connections between heavy quark thermalization and 
the quarkonium regeneration is left to the next secton. Wigner function is widely used for 
hadron production in the coalescence model. It gives the probability of $Q$ and $\bar Q$ 
combining into a bound state with relative distance $\vec r=\vec X_Q-\vec X_{\bar Q}$, 
and momentum $\vec q=\vec p_{Q}-\vec p_{\bar Q}$, see the function below~\cite{Greco:2003vf}
\begin{align}
\label{eq-wig}
f(r,q)=A_0 \exp(-{r^2\over \sigma^2(T)})\exp(-q^2\sigma^2(T))
\end{align}
$A_0$ is the normalization factor. Here we neglect the contributions 
of momentum carried by light partons in the heavy quark formation formation process, 
and use the $``b\bar b\rightarrow \Upsilon"$ to represent both $b+\bar b\rightarrow \Upsilon(1S) +g$ and 
$b+\bar b+\zeta\rightarrow \Upsilon(1S) +\zeta$. 
In realistic simulations, Only $0.01\%\sim 0.1\%$ of bottom quarks can form a bound state in the expanding 
QGP with a lifetime of $\sim10\ fm/c$ and an initial temperature of $\sim500$ MeV~\cite{Chen:2016mhl}. 
Most of them become open heavy-flavor hadrons with a light (anti-)quark. Considering the large ratio of 
$\sigma_{pp}^{b\bar b}/\sigma_{pp}^{\Upsilon(1S)}\sim 1000$ (much larger than the charm 
flavor $\sim 200$), this combination process may be important for the bottomonium 
nuclear modification factor $R_{AA}$. 
Bottomonium properties such as binding energy and shape of the wave function can be modified by 
the hot medium. This can affect the probability of heavy quark combination. We 
include hot medium effects on bottomonium properties by introducing 
the temperature dependence of Gaussian width 
$\sigma(T)$. It is connected with the bottomonium mean radius square at finite temperature by 
$\sigma(T)=\sqrt{8\langle r^2\rangle_\Upsilon(T)/3}$~\cite{Greco:2003vf}. 

In order to study the bottomonium regeneration in QGP, 
we put one $b$ and $\bar b$ in QGP and evolve each of them with Langevin equation to 
obtain the probability $W^{b\bar b\rightarrow \Upsilon}$ of their combination 
to regenerate a $\Upsilon(1S)$. 
The total yield of regenerated bottomonium within rapidity $\Delta y$ 
in nucleus-nucleus collisions is scaled by the number of bottom pairs as 
\begin{align}
N^{b\bar b\rightarrow \Upsilon }_{\rm{PbPb}}|_{\Delta y}=\sigma^{b\bar b}_{pp}|_{\Delta y}
N_{coll}\times 
W_{b\bar b\rightarrow \Upsilon }
\end{align}
The average number of $b\bar b$ pairs produced in central rapidity region is only $\sim 1$, 
the regenerated bottomonium yield is proportional to the number of $b\bar b$ pairs. 
The initial momentum and position of heavy quarks are randomly generated based 
on the distributions from PYTHIA 
simulations and Eq.(\ref{eq-den}) (both with modifications of the shadowing effect). 
At each time step, one 
can obtain their relative distance $r$ and relative momentum $q$, and their combination probability 
$P(r,q)=r^2q^2f(r,q)$. If the probability is larger than a random number between $0$ and $1$, 
then the formation process of bottomonium happens. Otherwise, they continue the evolutions 
with Langevin equation independently untill moving out of QGP (hadronization as open bottom hadrons). 
After $\Upsilon(1S)$ is regenerated inside QGP, it will decay due to parton inelastic scatterings 
and color 
screening from QGP. We supplement this part by the rate equation 
${dN_{\Upsilon}^{\rm rege}/dt} =-\Gamma^{\rm diss}_{\Upsilon}(T)N_\Upsilon$, 
where $\Gamma^{\rm diss}_{\Upsilon}$ is the decay rate from gluon 
and quasi-free dissociations. It is connected with Wigner function (see Eq.(\ref{eq-wig})). 

After one $\Upsilon(1S)$ is regenerated at the time $t_0$, 
the initial condition of Eq.(\ref{eq-upN}) 
becomes $N_{\Upsilon}(t_0)=1$, and $N_{\Upsilon}(t)$ 
decreases with time based on the rate equations, 
\begin{align}
\label{eq-upN}
N^{\rm rege}_\Upsilon(t_1+\Delta t) &= N_\Upsilon(t_1)e^{-\Gamma^{\rm diss}(T)\Delta t} \\
\label{eq-upR}
\vec R_{\Upsilon}(t_1 +\Delta t) &= \vec R_{\Upsilon}(t_1) +{\vec P_\Upsilon\over E_\Upsilon}\Delta t
\end{align}
Note that $N_{\Upsilon}(t\ge t_0)\le 1$ where $t_0$ is the time of $\Upsilon(1S)$ regeneration. 
$\vec P_\Upsilon\approx \vec p_b+\vec p_{\bar b}$ and $\vec R_\Upsilon$ are 
the momentum and coordinate of the center of the regenerated $\Upsilon(1S)$. 
From hydrodynamic equations, the local temperature of QGP is different at different coordinate $\vec R$ 
and time. Therefore, we update $\Upsilon(1S)$ position at each time step, and take a new local 
temperature $T$ of QGP at $\vec R_\Upsilon(t_1+\Delta t)$ 
to recalculate the decay 
rate $\Gamma^{\rm diss}$ for $\Upsilon(1S)$ evolution at the next time step. 
We continue Eq.(\ref{eq-upN}-\ref{eq-upR}) from the time $t_0$ of $\Upsilon(1S)$ regeneration  
untill it moves out of QGP (where 
local temperature is smaller than the critical temperature $T_c$). 
Doing sufficient events, $N_{b\bar b}^{\rm events}$, of putting one $b$ 
and $\bar b$ in the expanding QGP, and sum $\Upsilon(1S)$ final 
yields to be $N_{\Upsilon}^{\rm rege}(\rm tot)$, one can obtain 
the $\Upsilon(1S)$ regeneration probability from $b$ 
and $\bar b$ evolutions in QGP, 
$W_{b\bar b\rightarrow \Upsilon }
=N_{\Upsilon}^{\rm rege}(\rm tot)/N_{b\bar b}^{\rm events}$. 

\subsection{Bottomonium Dissociation and Regeneration Rates at Finite Temperature}
\begin{figure}[!t]
\centering
\includegraphics[width=0.4\textwidth]{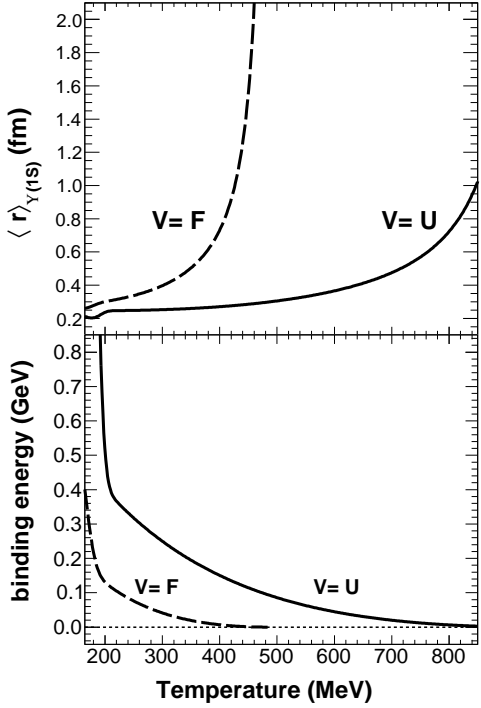}
\caption{
Upper panel: mean radius of $\Upsilon(1S)$ at finite temperature with the heavy quark potential 
to be two limits: free energy V=F (dashed line) and internal energy V=U (solid line). 
Lower panel: binding energy of $\Upsilon(1S)$ with V=F and V=U respectively. 
}
\hspace{-0.1mm}
\label{fig-radius}
\end{figure}

QGP color screening reduces the binding energy of bottomonium, and increases its decay rate at 
finite temperature. With heavy quark potential to be $V_{b\bar b}=F$ or 
$V_{b\bar b}=U$~\cite{Liu:2010ej}, One can obtain the 
$\Upsilon(1S)$ binding energy from solving time-independent radial 
Schr\"odinger equation (with $\hbar =c=1$)
\begin{align}
\label{eq-sch}
[-{1\over 2m_\mu}{\partial ^2\over \partial r^2} +V_{b\bar b}(r,T)]\psi(r) 
= \mathcal{E}_\Upsilon(T) \psi(r)
\end{align}
Here $m_\mu=m_b/2$ is the the reduced mass in the center of $\Upsilon$ mass frame. 
$\mathcal{E}_\Upsilon(T)$ and $\psi(r)$ are 
the binding energy and radial wave function of a $\Upsilon$ eigenstate. 
The mean radius and binding energy (see Fig.\ref{fig-radius}) 
obtained consistently from Eq.(\ref{eq-sch}) will be used in the 
Wigner function for bottomonium regeneration and the parton dissociations, respectively. 

\begin{figure}[!t]
\centering
\includegraphics[width=0.4\textwidth]{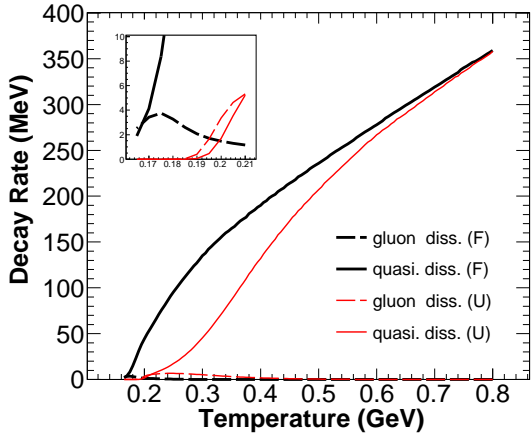}
\caption{(Color online) Decay rates of $\Upsilon(1S)$ as a 
function of temperature. Dashed and solid lines are 
the decay rates from gluon and quasi-free dissociations respectively. Thick black and thin red lines 
are with heavy quark potential to be the free energy $F$ and internal energy $U$.  
}
\hspace{-0.1mm}
\label{fig-decay}
\end{figure}

The decay rates from gluon dissociation and 
quasi-free dissociation ~\cite{Grandchamp:2003uw,Zhao:2010nk} with temperature dependent 
binding energy are plotted in Fig.\ref{fig-decay}. 
When the $\Upsilon(1S)$ is strongly bound, gluon dissociation dominates 
the $\Upsilon(1S)$ decay rate, such as at the temperature region of $T<0.2$ GeV with $V_{b\bar b}=U$ 
(see red dashed and solid lines).
However at a high temperature like $T=300$ MeV, strong color screening effect reduces 
the $\Upsilon(1S)$ binding 
energy to be around $0.25$ GeV with V=U and $0.04$ GeV with V=F, which are 
far below the vacuum value $\sim1.1$ GeV~\cite{Laine:2006ns}. 
This makes quasi-free dissociation dominates the 
decay rate. We include both contributions on the hot medium 
suppression of regenerated $\Upsilon(1S)$ at the entire temperature region $T>T_c$. 

\section{hydrodynamic model}
We employ the (2+1) dimensional ideal hydrodynamics to simulate strong expansion of finite sized 
QGP produced in ultrarelativistic heavy ion collisions, 
\begin{align}
\label{Eqhydro1}
\partial _\mu T^{\mu\nu}=0 
\end{align}
Here $T^{\mu\nu}=(e+p)u^\mu u^\nu-g^{\mu \nu} p$ is the energy-momentum tensor, and 
($e,p,u^\mu$) are the energy density, pressure and four-velocity of fluid cells.  
The equation of state of the deconfined medium is taken as 
an ideal gas of massless ($u,d$) quarks, 
150 MeV massed $s$ quarks and gluons~\cite{Sollfrank:1996hd}. Hadron
phase is an ideal gas of all known hadrons and resonances
with mass up to 2 GeV~\cite{Hagiwara:2002fs}. From the scaling of 
initial temperature at $\sqrt{s_{NN}}=2.76$ 
TeV Pb-Pb collisions which is $T_0 =485$ MeV, 
we set the initial maximum temperature at $\sqrt{s_{NN}}=5.02$ TeV 
to be $T_0=510$ MeV~\cite{Chang:2015hqa}.  
Based on hydrodynamic model studies, light hadron spectra 
at RHIC 200 GeV Au-Au and LHC 2.76 TeV Pb-Pb collisions 
indicate a same time scale of QGP reaching local equilibrium 
$\tau_0^{\rm 200 GeV} \approx \tau_0^{\rm 2.76 TeV}\approx 0.6$ fm/c~\cite{Shen:2012vn,Hirano:2001eu}. 
Therefore, we still take the 
same value of $\tau_0=0.6$ fm/c at 5.02 TeV Pb-Pb collisions due to its weak dependence 
on colliding energy. The transverse expansion of QGP controlled by Eq.(\ref{Eqhydro1}) 
starts from $\tau_0$.

\section{inputs of bottom flavor}
The bottomonium regeneration requires the number of bottom quarks in nucleus-nucleus collisions. 
We determine this by using the production cross section in 
$pp$ collisions and binary collision scaling in Pb-Pb collisions, 
$N_{\rm PbPb}^{b\bar b}=\sigma_{pp}^{b\bar b}N_{coll}(b)$. Lack of experimental data about 
$\sigma_{pp}^{b\bar b}$ at 5.02 TeV $pp$ collisions, 
we extract its value by the linear interpolation between the cross sections at central rapidity at 
1.96 TeV and 2.76 TeV collisions. 
At 1.96 TeV pp collisions, CDF collaboration published the cross 
section of b-hadrons integrated over all transverse momenta 
in the rapidity $|y|<0.6$ to be $17.6\pm0.4(stat)^{+2.5}_{-2.3}(syst)\ \mu b$~\cite{Acosta:2004yw}. 
With this we extract the central value of the differential cross section 
to be $d\sigma_{pp}^{b\bar b}/dy=14.7\ \mu b$. Combined 
with the $d\sigma_{pp}^{b\bar b}/dy =23.28\pm 2.70(stat)^{+8.92}_{-8.70}(syst)\ 
\mu b$ in the central rapidity 
at 2.76 TeV~\cite{Abelev:2014hla} , we obtain the differential cross section 
$d\sigma_{pp}^{b\bar b}/dy=47.5\ \mu b$ in the central rapidity at 5.02 TeV pp collisions. 
Our purpose is to study the contribution of regeneration component in the 
experimentally measured inclusive $\Upsilon(1S)$ yields, 
presented as the nuclear modification factor~\cite{Chen:2015iga},
\begin{align}
\label{eq-RAA}
R_{AA}^{\rm inclu}(\Upsilon(1S)) &= 
{N_{AA}^{\rm prim}(\Upsilon(1S)) 
+N_{AA}^{b\bar b\rightarrow \Upsilon(1S) }\over {d\sigma_{pp}^{\Upsilon(1S)}\over dy} \Delta y
\cdot N_{coll}(b)}  \nonumber \\ 
&= R_{AA}^{\rm prim} + R_{AA}^{\rm rege} \\
\label{eq-regeN}
N_{AA}^{b\bar b\rightarrow \Upsilon(1S) } &=W^{\rm Lan+Wigner}_{b\bar b\rightarrow \Upsilon(1S) } 
({d\sigma_{pp}^{b\bar b}\over dy}\Delta y\cdot N_{coll})
\end{align}
where $N_{AA}^{\rm prim}$ in the numerator of Eq.(\ref{eq-RAA}) 
represents the $\Upsilon(1S)$ primordial production including 
direct production and decay contributions from excited states (1P,2P,2S,3S). 
The second term $N_{AA}^{b\bar b\rightarrow \Upsilon(1S) }$ is for the 
$new$ bottomonium from $b$ and $\bar b$ combination during QGP evolutions, 
which is closely connected with the bottom quark diffusions 
in the expanding QGP and so our main interest in this work.  
$W^{\rm Lan+Wigner}_{b\bar b\rightarrow \Upsilon(1S) }$ is the 
probability of one $b$ and $\bar b$ quark 
combine into a $\Upsilon(1S)$. 
$d\sigma_{pp}^{\Upsilon(1S)}/dy$ in the denominator of Eq.(\ref{eq-RAA}) is 
the $\Upsilon(1S)$ inclusive cross section. 
With the differential cross sections $d\sigma_{pp}^{\Upsilon(1S)}/dy=27\pm1.5\ nb$ 
at 1.8 TeV~\cite{Acosta:2001gv} and $d\sigma_{pp}^{\Upsilon(1S)}/dy=80\pm 9$ nb 
at 7 TeV from CMS Collaboration~\cite{Khachatryan:2010zg} in the cetral rapidity of 
pp collisions, we extract the central value of inclusive cross section to 
be $d\sigma_{pp}^{\Upsilon(1S)}/dy=59.8$ nb at 5.02 TeV, which gives the  
ratio of $N_{pp}^{\Upsilon(1S)}/N^{b\bar b}_{pp}$ in central rapidity 
to be $0.13\%$, close to the typical order $0.1\%$ of hidden- to open-bottom state ratio 
in elementary hadronic collisions~\cite{Emerick:2011xu}. 

The coupling strength between bottom quarks and QGP is indicated by the drag coefficient 
in Langevin equation. 
The spatial diffusion coefficient is taken to be 
$D(2\pi T)=C$~\cite{Cao:2015cba}. Different values of 
$C$ will be employed to study the effects of bottom quark thermalization 
on bottomonium regeneration.

\section{Bottomonium Continuous Regeneration in $\rm Pb-Pb$ collisions}
In the previous work~\cite{Chen:2016mhl}, we employ the Langevin equation plus Wigner function to 
calculate $J/\psi$ and $\psi(2S)$ regeneration from dynamical evolutions of (anti-)charm quarks, 
with an assumption that they are produced at 
each certain temperature, without the following suppression from hot medium after their 
regeneration. 
In this work, we improve our approach of ``Langevin equation +Wigner function'' by 
considering hot medium modifications on quarkonium properties, which change quarkonium regeneration 
and dissociation rates through the mean radius $\langle r\rangle_{\Upsilon}(T)$ and 
binding energy $\mathcal{E}_\Upsilon(T)$ of quarkonium. 

In the realistic simulations, we generate one $b$ and $\bar b$ 
randomly in the coordinate and momentum space based on the 
probability distributions given in previous sections. Then we evolve them separately with two 
individual Langevin equations, and check if they can form a $\Upsilon(1S)$ at 
each time step. In central collisions, $b$ and $\bar b$ are easier  
to lose energy and meet each other to form a new $\Upsilon(1S)$. Smaller value of the parameter $C$ 
indicate a stronger coupling strength between bottom quarks and QGP, 
which results in larger regeneration rate $W_{b\bar b\rightarrow \Upsilon }^{\rm Lan+Wigner}$. 

\begin{figure}[!t]
\centering
\includegraphics[width=0.4\textwidth]{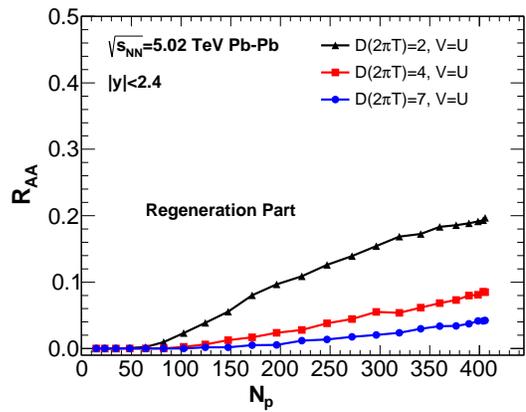}
\caption{
(Color online) Nuclear modification factor of 
regenerated $\Upsilon(1S)$ from $b$ and $\bar b$ quark 
combination as a function of number of participants $N_p$ in central rapidity region $|y|<2.4$ at 
5.02 TeV Pb-Pb collisions. The heavy quark potential is taken as internal energy V=U 
(used to determine the mean radius and binding energy of $\Upsilon(1S)$). Lines with triangle, square 
and circle markers are for diffusion coefficient $D(2\pi T)=2,4,7$ respectively.
Note that
these $D$ values satisfy the relation of $1\lesssim D(2\pi T)\lesssim 7$ from pQCD and
Lattice calculations~\cite{CaronHuot:2007gq,Kovtun:2003wp,Ding:2012sp}.
}
\hspace{-0.1mm}
\label{fig-regRAA}
\end{figure}

Now we do the full calculations of bottomonium regeneration in 
heavy ion collisions, and give the regeneration part (see Eq.(\ref{eq-RAA}-\ref{eq-regeN})) 
of inclusive $R_{AA}^{\Upsilon(1S)}$ in Fig.\ref{fig-regRAA}. 
In central collisions where QGP temperature is high, 
both number of $b\bar b$ pairs and the combination probability of $b$ and $\bar b$ quarks 
become large in Pb-Pb collisions. 
This makes $R_{AA}^{\Upsilon(1S)}$ increase with $N_p$. The slope of $R_{AA}^{\Upsilon(1S)}(N_p)$ is 
larger than the slopes of $N_{\rm PbPb}^{b\bar b}(N_p)$ 
and $W_{b\bar b\rightarrow \Upsilon }^{\rm Lan+Wigner}(N_p)$.  
From the nuclear modification factor of B-hadrons, the situation of $D(2\pi T)=4$ seems better to 
describe the heavy quark energy loss in $\sqrt{s_{NN}}=5.02$ TeV Pb-Pb collisions~\cite{Cao:2015cba}.
In the most central collisions, the regeneration contributes 
to the inclusive $R_{AA}^{\Upsilon(1S)}$ 
with around $(10\sim20)\%$ (V=U) and $(5\sim10)\%$ (V=F). 
But it is only $\sim4\%$ at minimum bias centrality (with $N_p\approx 200$). 
Note that transport model calculations gave the regeneration $R_{AA}$ to be $\sim 8\%$ 
in semi-central ($N_p\approx 200$) 
and most cenral Pb-Pb collisions in the rapidity $|y|<2.4$ 
at 2.76 TeV~\cite{Zhao:2012gc,Emerick:2011xu}. 

One way to testify the contribution of $\Upsilon(1S)$ regeneration 
in heavy ion collisions is to study the 
rapidity dependence of the $p_T-$integrated $R_{AA}^{\Upsilon(1S)}(y)$. 
The bottom quark differential cross section decreases with rapidity, 
which will suppress the $\Upsilon(1S)$ 
regeneration at forward rapidity. For charmonium, the decreasing 
tendency of $R_{AA}^{J/\psi}(y)$ with rapidity is very strong and explained well by the 
regeneration mechanism~\cite{Chen:2015iga}. 
As charmonium regeneration mainly dominates at the low $p_T$ bins and 
drops to zero at middle and high $p_T$ bins. Therefore, $R_{AA}^{J/\psi}$ shows strong decreasing 
tendency with rapidity at $p_T>0$ (where regeneration dominates) and almost no rapidity 
dependence at $p_T\gtrsim4$ GeV/c. Considering the fraction of 
$\Upsilon(1S)$ regeneration is only $10\sim 15\%$ in its inclusive $R_{AA}$ at the impact parameter b=0, 
we expect a weaker rapidity dependence of $R_{AA}^{\Upsilon(1S)}(y)$. 
In the minimum bias centrality, 
This tendency should be much weaker and $R_{AA}^{\Upsilon(1S)}(y)$ almost shows a flat feature, 
just as experimental data 
shows~\cite{Abelev:2014nua,CMS:2017ucd}. 

\begin{figure}[!t]
\centering
\includegraphics[width=0.4\textwidth]{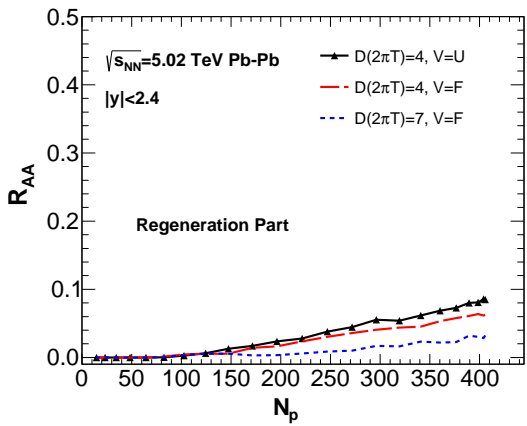}
\caption{
(Color online) 
Nuclear modification factor of regenerated $\Upsilon(1S)$ from $b$ and $\bar b$ quark 
combination as a function of number of participants $N_p$ in central rapidity region $|y|<2.4$ at 
5.02 TeV Pb-Pb collisions. The heavy quark potential is taken as V=U and V=F. 
}
\hspace{-0.1mm}
\label{fig-UFRAA}
\end{figure}

We also did the calculations in $\Upsilon(1S)$ weak binding 
scenario (V=F), see Fig.\ref{fig-UFRAA}. With weak binding (V=F), 
$\Upsilon(1S)$ can only be regenerated at $T<400$ MeV where its binding energy is non-zero. 
Also, the dissociation rate of regenerated $\Upsilon(1S)$ is larger for V=F compared with V=U, which 
makes regenerated $\Upsilon(1S)$ easier to be dissociated. 
$\Upsilon(1S)$ regeneration with V=F is smaller (see Fig.\ref{fig-UFRAA}). This is also 
consistent with transport model calculations. The $\Upsilon(1S)$ $R_{AA}$ from regeneration 
contribution is around $6\%$ (with $D(2\pi T)=4$) and $3\%$ (with $D(2\pi T)=7$) in the most central 
collisions. In all centralities, its contribution is smaller than the value of $V=U$, 
see Fig.\ref{fig-UFRAA}.  

\section{Summary}
We employ the Langevin equation to describe the diffusion of bottom quarks, and Wigner function for 
the $b$ and $\bar b$ quarks to regenerate $\Upsilon(1S)$s during the 
QGP evolutions. After the regeneration 
of a $\Upsilon(1S)$, it also suffers the color screening and parton inelastic scatterings from QGP. 
The dissociation and regeneration rates of $\Upsilon(1S)$ are connected with each other by the temperature 
dependent binding energy $\mathcal{E}_{\Upsilon}(T)$ and mean radius $\langle r\rangle_\Upsilon(T)$, 
which can be obtained simultaneously from Schr\"odinger equation 
with the color screened heavy quark potential extracted from Lattice calculations. 
Supplement the cold nuclear matter effects, we give the full calculations of $\Upsilon(1S)$ regeneration 
in Pb-Pb collisions at $\sqrt{s_{NN}}=5.02$ TeV. With different drag coefficient in Langevin equation, 
the energy loss of $b$ and $\bar b$ quarks is different. 
Strong coupling between bottom quarks 
and QGP can make $b$ and $\bar b$ lose more energy, and  
increases the probability of their formation into a bound state. In the 
scenario of V=U (V=F), we obtain the 
$\Upsilon(1S)$ regeneration $R_{AA}^{\rm rege}$ to be $0.1\sim 0.2$ ($0.06\sim 0.1$) 
in the most central collisions, but 
negligible in minimum bias centrality. With realistic evolutions of bottom quarks and 
their hadronization process, we study the regeneration contribution to the $\Upsilon(1S)$ 
nuclear modification factor $R_{AA}$ measured in experiments, and also build 
the connection between bottom quark energy loss and bottomonium 
regeneration in this work.

\vspace{0.2cm}
\appendix {\bf Acknowledgement: } Baoyi Chen thanks X. Du for helpful discussions. We 
acknowledge that TAMU group is also studying bottomonium regeneration based on Langevin equation 
independently. Please see their proceeding of Quark Matter 2017~\cite{Du:2017hss}. 
This work is supported by the NSFC under Grant No. 11547043.


\end{document}